\newcommand{\PRL}[3]{Phys.\ Rev.\ Lett.\ {\bf #1},\ #2 (#3)}

\newcommand{\NAT}[3]{Nature\ {\bf #1},\ #2 (#3)}
\newcommand{\SC}[3]{Science\ {\bf #1},\ #2 (#3)}

\newcommand{\PRA}[3]{Phys.\ Rev.\ A\ {\bf #1},\ #2 (#3)}

\newcommand{\PRE}[3]{Phys.\ Rev.\ E\ {\bf #1},\ #2 (#3)}

\newcommand{\JPB}[3]{J.\ Phys.\ B:\ At.\ Mol.\ Opt.\ Phys.\ {\bf #1},\ #2 (#3)}

\documentclass[showpacs,preprintnumbers,amsmath,twocolumn,amssymb,pra,superscriptaddress]{revtex4}
\usepackage{latexsym}
 \usepackage{amsmath}
 \usepackage{amsfonts}
 \usepackage{graphicx}
 \usepackage{amssymb}
 \usepackage{dsfont}
 \usepackage{subfigure}

\begin{document}

\title{New cross-phase modulated localized solitons in coupled atomic-molecular
BEC}

\author{Challenger Mishra}
\email{challenger@iiserkol.ac.in}
\affiliation{Indian Institute of Science Education and Research Kolkata, Mohanpur
741252, India}

\author{Priyam Das}
\email{pprasanta@iiserkol.ac.in}
\affiliation{Indian Institute of Science Education and Research Kolkata, Mohanpur
741252, India}

\author{Krishna Rai Dastidar}
\email{spkrd@iacs.res.in}
\affiliation{Indian Association for the Cultivation of Science, Jadavpur, Kokata
- 700032, India}

\author{P. K. Panigrahi}
\email{pprasanta@iiserkol.ac.in}
\affiliation{Indian Institute of Science Education and Research Kolkata, Mohanpur
741252, India}


\begin{abstract}
The interacting atom-molecule BEC (AMBEC) dynamics is investigated in the mean field approach. The presence of atom-atom, atom-molecule and molecule-molecule interactions, coupled with a characteristically different interaction representing atom-molecule interconversion, endows this system with nonlinearities, which differ significantly from the standard Gross-Pitaevskii (GP) equation. Exact localized solutions are found to belong to two distinct classes. The first ones are analogous to the soliton solutions of the weakly coupled GP equation, whereas the second non-equivalent class is related to the solitons of the strongly coupled BEC. Distinct parameter domains characterize these solitons, some of which are analogous to the complex profile Bloch solitons in magnetic systems. These localized solutions are found to represent a variety of phenomena, which include co-existence of both atom-molecule complex and miscible-immiscible phases. Numerical stability is explicitly checked, as also the stability analysis based on the study of quantum fluctuations around our solutions. We also find out the domain of modulation instability in this system.\end{abstract}
\maketitle

\section{Introduction}

Molecular BECs have been experimentally realized in recent times \cite{wynar,gerton,donley,mark,winkler,danzl}. Co-existence and inter conversion of atomic and molecular Bose-Einstein
condensates (BECs) have been observed experimentally. Raman photoassociation is an important process by which the molecular species in an AMBEC can be formed. This was investigated theoretically in Refs \cite{jjavanian,drum,javanian,drummond1}. The mean field
description of the same involves generalization of the Gross-Pitaevskii
equation to take into account atom-molecular two-body scattering,
as well as the atom-molecule inter-conversion. There are theoretical predictions that the atomic and molecular species can show distinct collective oscillations in an AMBEC \cite{heinzen,hope,oliveira}. In cigar-shaped BEC,
this rich dynamical system paves the way for observation of novel
solitons and nonlinear periodic waves, akin to the fundamental dark
and bright solitons of the atomic BEC, in the repulsive and attractive
regimes \cite{burger,deng,khaykovich,strecker,khawaja,cornish}. The fact that GP equation in one dimension is the integrable
nonlinear Schr\"odinger equation, which admits soliton solutions, has
led to considerable theoretical and experimental investigations of
the cigar-shaped BEC. Dark solitons, bright solitons and soliton trains
have been experimentally observed \cite{burger,deng,khaykovich,strecker}. Novel instability mechanisms have
been proposed for the break up of bright soliton and formation of
soliton trains, since modulation instability has not been adequate
in explaining the same \cite{konotop,strecker}. The two-component BEC (TBEC) is, in general non-integrable, having close connection with the integrable
Manakov system \cite{manakov}. The soliton solutions and their structure and stability
has been extensively studied for this system, both analytically and
numerically. The mean field equations describing the atom-molecule
BEC complex, has close similarity, with both weakly and strongly coupled
atomic BEC. The two-body atom-atom, molecule-molecule and atom-molecule
scattering terms are analogous to cubic nonlinearity of the standard
GP equation, whereas the quadratic nonlinear terms arising from atom-molecule
conversion is identical to the nonlinear interaction term in strongly-coupled
BEC, in one-dimension \cite{salasnich}. This dual structure of the interaction terms,
provides a novel form of cross-phase modulation, not possible in the
conventional TBEC case.

\section{The mean field description of AMBEC}

In the absence of the trap, the mean-field dynamics of the cigar-shaped AMBEC complex is governed by the mean-field equations \cite{olessacha}:


\begin{align}
i\frac{\partial\phi_{a}}{\partial t}&=\left[-\frac{1}{2}\frac{\partial^{2}}{\partial x^{2}}+\lambda_{a}N|\phi_{a}|^{2}+\lambda_{am}N|\phi_{m}|^{2}\right]\phi_{a}\nonumber\\
&+\alpha\sqrt{2N}\phi_{m}\phi_{a}^{*}\ ,\\
i\frac{\partial\phi_{m}}{\partial t}&=\left[-\frac{1}{4}\frac{\partial^{2}}{\partial x^{2}}+\epsilon+\lambda_{m}N|\phi_{m}|^{2}+\lambda_{am}N|\phi_{a}|^{2}\right]\phi_{m}\nonumber\\
&+\alpha\sqrt{\frac{N}{2}}\phi_{a}^{2} \ . 
\end{align}


where N is the total number of atoms in the system. Here, $\epsilon$
is the binding energy and the terms with coefficients $\lambda_{a}$,
$\lambda_{m}$ and $\lambda_{am}$ denote the effect of atom-atom,
molecule-molecule and atom-molecule collisions, respectively. The interaction
involving $\alpha$ denote the conversion of atoms to molecule and
vice-versa. In comparison, the mean-field GP equations in the weak and strong
coupling sectors are given by \cite{salasnich},

\begin{equation}
i\hbar\partial_{t}\phi=\big[-\frac{\hbar^{2}}{2m}\partial_{z}^{2}+2\hbar\omega_{\perp}a\big(|\phi|^{2}-\sigma_{0}\big)\big]\phi
\end{equation}

under the condition $2aN|\phi|^{2}\ll1$ and

\begin{equation}
i\hbar\partial_{t}\phi=\big[-\frac{\hbar^{2}}{2m}\partial_{z}^{2}+2\hbar\omega_{\perp}a^{1/2}\big(|\phi|-\sigma_{0}^{1/2}\big)\big]\phi 
\end{equation}

under the condition $2aN|\phi|^{2}\gg1$ respectively. $\omega_\perp$ is the trapping frequency in the radial direction. N is the number of atoms in the condensate, $a$ is the scattering length and $\sigma_{0}$ is the equilibrium density of atoms, far away from
the axis. These two arise from the non-polynomial interaction
term  $\frac{|\phi|^2}{\sqrt{1+2a\mathrm{N}|\phi|^2}}\phi$, when a dimensional reduction of the 3D GP equation is carried out for the cigar shaped BEC. In the weak coupling case, one can neglect
the $|\phi|^{2}$ term in the denominator, yielding the familiar
cubic nonlinearity, whereas for the strong coupling case one gets
$\frac{|\phi|^{2}}{|\phi|}\phi=|\phi|\phi$. The structure of the
solutions are quite different in these two sectors. The dark and bright
solitons are of the type A $\tanh\left[(x-ut)\frac{\cos\theta}{\xi}\right]$
and B $\mathrm{sech}\left[(x-ut)\frac{\cos\theta}{\xi}\right]$ for the weakly
coupled case, whereas in the later case, one finds the solutions are
of the type A $\mathrm{sech}^{2}\left[(x-ut)\frac{\cos\theta}{\xi}\right]$
+ constant  \cite{kumarpp}.

In the following, we highlight the above mentioned similarities
between atom-molecular BEC and weak-strong coupled atomic BEC. The
different nature of the localized modes in the two different regimes,
is pointed out. Subsequently, we exhibit the exact solutions of this
dynamical system, which realize a new cross-phase modulation, arising
due to atom-molecule interaction. We check the stability of our solution numerically as well as based on the study of quantum fluctuations around our solutions. We also find out the domain of modulation instability. We then conclude with directions for future investigation in this rich dynamical system.

\section{Soliton solutions}

\subsection{Sech-Tanh complex soliton pair}

For identifying exact solitonic solutions to the equations [1] and [2], we start with the following ansatz for the mean fields:

\begin{align}
\phi_{a}(x,t) &= a\cos\theta \textrm{sech}\left[(x-ut)\frac{\cos(\theta)}{\xi}\right]\exp[i\ (px-\Omega t)],\\
\intertext{and}\phi_{m}(x,t) &=  ib\left[\cos\theta \textrm{tanh}\left[(x-ut)\frac{\cos(\theta)}{\xi}\right]+i\sin(\theta)\right] \nonumber\\
&\textrm{x}\exp[2i\ (px-\Omega t)]\end{align}

This ansatz solution represents the physical scenario, where the asymptotic condensate density of atoms vanishes, and that of molecules reaches a constant value. The equations of motion yield the following consistency relations:

\begin{align}
0&=\big[\frac{1}{2}p^{2}-\Omega+\frac{1}{2\xi^{2}}\cos^{2}\theta+(\lambda_{a}a^{2}\cos^{2}\theta+
\lambda_{am}b^{2}\sin^{2}\theta)N\nonumber \\
&-\alpha\sqrt{2N}b\sin\theta\big]\ ,\\
0&=\left[\frac{u-p}{\xi}-\alpha\sqrt{2N}b\right]\ ,\\
0&=\left[\frac{1}{\xi^{2}}+\left(\lambda_{a}a^{2}-\lambda_{am}b^{2}\right)N\right]\ ,\\
0&=\left[\frac{u-p}{\xi}-\alpha\sqrt{\frac{N}{2}}\frac{a^{2}}{b}+\left(\lambda_{am}a^{2}-\lambda_{m}b^{2}\right)N\sin\theta\right]\ ,\\
0&=\left[b^{2}N\lambda_{m}+\epsilon+p^{2}-2\Omega\right]\ ,\\
\intertext{and}0&=\left[\frac{1}{2\xi^{2}}+\left(\lambda_{am}a^{2}-\lambda_{m}b^{2}\right)N\right]\ .
\end{align}

A lengthy but straightforward calculation leads to the following solutions:

\begin{align}
b^{2} & =  \frac{\gamma_{3}}{N\gamma_{4}}\ ,\\
\sin^{2}\theta & =  \frac{\alpha^{2}\gamma_4}{2\gamma_3}(\gamma_{2}/\gamma_{1})^{2}\ ,\\
\xi^{2} & =  \frac{(\lambda_{a}-2\lambda_{am})\gamma_4}{2\gamma_1\gamma_{3}}\ ,\\
\Omega-\frac{1}{2}p^{2} & =  (\lambda_{m}\frac{\gamma_3}{\gamma_4}+\epsilon)/2\  ,\\
a^{2} & =  \frac{(\lambda_{am}-2\lambda_{m})}{(\lambda_{a}-2\lambda_{am})}\frac{\gamma_3}{N\gamma_{4}}\ ,\\ 
\intertext{and}(u-p)^{2} & =  \frac{\lambda_a-2\lambda_{am}}{\gamma_1}\alpha^{2}\ ;
\end{align}

where
\begin{align*}
\gamma_{1}&=\left(\lambda_{a}\lambda_{m}-\lambda_{am}^{2}\right)\ ,\\  
\gamma_{2}&=\left(2\lambda_{a}+2\lambda_{m}-5\lambda_{am}\right)\ ,\\ 
\gamma_{3}&=\alpha^{2}\left[\frac{\gamma_{2}}{\gamma_{1}}-\frac{\gamma_{2}^{2}}{2\gamma_{1}\left(\lambda_{a}-2\lambda_{am}\right)}\right]+\frac{\epsilon}{2}\ ,\\
\intertext{and}\gamma_{4}&=\lambda_{am}-\frac{\lambda_{m}}{2}-\frac{\gamma_{1}}{\left(\lambda_{a}-2\lambda_{am}\right)}\ . 
\end{align*}

One can now put any valid value of momentum $p$ to the above equations and obtain
the remaining variables exactly in terms of the parameters of the
equations (1) and (2). From the above equations, it is easy to see, for $p=0$, that the velocity of the solitons and the depth of the grey soliton, are completely fixed by the parameters in the mean field equations (1) and (2). Hence, in order to obtain grey solitons of varying depths, one needs to tune the parameters in the theory.

\begin{figure}[!t]
\noindent \centering{}
\includegraphics[scale=0.4]{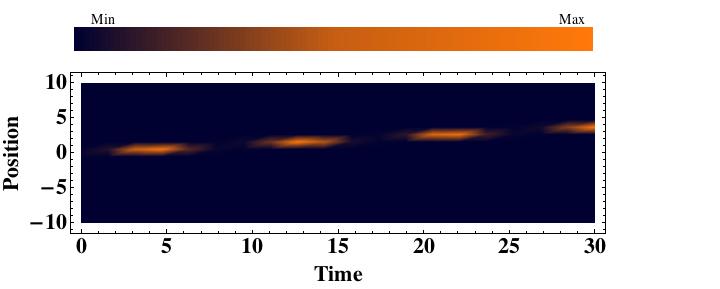}
\includegraphics[scale=0.4]{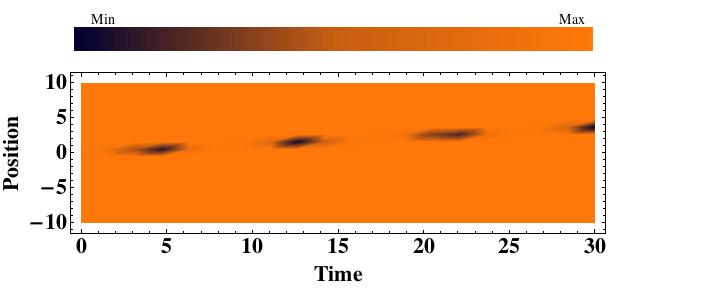}
\caption{ (top) Bright soliton (5) and (bottom) grey soliton (6) propagation snapshots at different times. $\lambda_a = 3\mathrm{x}10^4, \lambda_m = 10^5, \lambda_{am} = 80.5\mathrm{x}10^3, \alpha = 20, \mathrm{N} = 100, \epsilon = 0, a = 0.000132, b = 0.000138, \theta = 
 0.4336, \xi = 3.147917, \Omega = 0.094963\  \textrm{and} \  u = 0.122704$.}
\end{figure}

\subsection{New class of Soliton solutions}

Due to the nature of the interconversion term in equations (1) and (2), only very special classes of solutions are allowed in the system. Another class of solutions for the AMBEC is presented, which resemble the solutions in the strongly coupled BECs \cite{kumarpp}. This is a special class of solution since the atomic and molecular phases have the same density profiles at all times and therefore the atoms and the molecules are always in a miscible phase. The soliton profiles are given by,

\begin{align}
\phi_{a}(x,t) & =  a\left[1-\sigma\tanh^{2}\left[\gamma(x-ut)\right]\right]\exp[i\ (px-\Omega t)]\ ,\\
\intertext{and}
\phi_{m}(x,t) & =  a\left[1-\sigma\tanh^{2}\left[\gamma(x-ut)\right]\right]\exp[2i\ (px-\Omega t)]
\end{align}

The consistency conditions allow only two discrete values of the parameter $\sigma$:

\begin{align}
\sigma & =  1, 3\ ,\\
\lambda_{a} & =  \lambda_{m}=-\lambda_{am}\ ,\\
a & =  \frac{\epsilon}{\sqrt{2N}\alpha}\ ,-\frac{\epsilon}{3\sqrt{2N}\alpha}\ ,\\
\gamma & =  \sqrt{-\epsilon/3} ,  \sqrt{\epsilon/3}\ ,\\
\intertext{and}
\Omega-p^{2}/2 & =  2\epsilon/3
\end{align}

\begin{figure}[!t]
\noindent \centering{}
\includegraphics[scale=0.45]{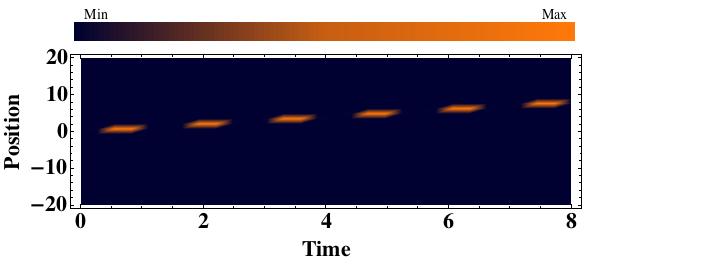}
\includegraphics[scale=0.45]{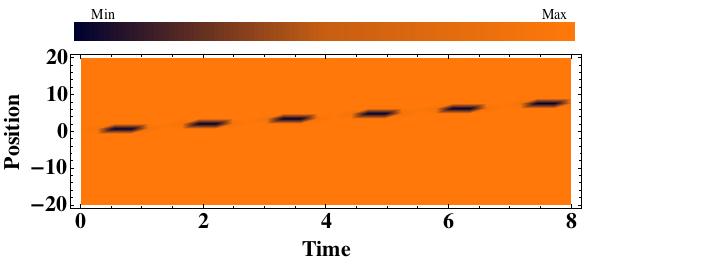}
\caption{ Depiction of $|\phi_a|^2$ for the second class of solution for $\beta=1$ (top) and for $\beta=3$ (bottom). Parameters used are p = 1,  $\epsilon = 100, \mathrm{N} = 100 \ \textrm{and} \  \alpha = 41$. $|\phi_m|^2$ has the same profile. The figure represents snapshots at discrete times. }
\end{figure}

\section{Stability under Quantum fluctuations}
We now investigate the stability of the obtained solutions, using the method of C. K. Law et. al. \cite{eberlystability}. This analysis revealed the regime of coupling parameters in the theory, in which, the ground state solutions to the coupled NLSE, were stable under vacuum fluctuations.  The above authors identified an eigenvalue, associated with the system as the determiner of condensate stability. It plays the same role as the sign of scattering length in a single species condensate. We will carry out a similar analysis here to find out the parameter domains, where the complex bright-grey pair of soliton solution is stable. To perform the analysis, one starts with the second quantized grand canonical Hamiltonian of the atom-molecule BEC:

\begin{align*}
\hat{H}&=\int{\rm d}^{3}r\Big(\hat{\psi}_{a}^{\dagger}\Big[-\frac{\hbar^{2}}{2m}\nabla^{2}+U_{a}(\vec{r})+\frac{\lambda_{a}}{2}\hat{\psi}_{a}^{\dagger}\hat{\psi}_{a}\Big]\hat{\psi}_{a}\\
&+\hat{\psi}_{m}^{\dagger}\Big[-\frac{\hbar^{2}}{4m}\nabla^{2}+U_{m}(\vec{r})+{\cal E}+\frac{\lambda_{m}}{2}\hat{\psi}_{m}^{\dagger}\hat{\psi}_{m}\Big]\hat{\psi}_{m}\\
&+\lambda_{am}\hat{\psi}_{a}^{\dagger}\hat{\psi}_{a}\hat{\psi}_{m}^{\dagger}\hat{\psi}_{m}
+\frac{\alpha}{\sqrt{2}}\big[\hat{\psi}_{m}^{\dagger}\hat{\psi}_{a}\hat{\psi}_{a}+\hat{\psi}_{m}\hat{\psi}_{a}^{\dagger}\hat{\psi}_{a}^{\dagger}\big]\Big)
\end{align*}

At the temperature $T=0\ K$, we linearize this Hamiltonian by assuming 

\begin{eqnarray}
\hat{\psi}_a&=&\phi_a + \psi_a\\
\hat{\psi}_m&=&\phi_m + \psi_m
\end{eqnarray}

where $\phi_a$ and $\phi_m$ are solutions to the coupled GP equations. The bright-grey pair solution, (5) and (6), that was  obtained for AMBEC and was tested for its stability. The fluctuation part of $\hat{\psi}_a(m)$ is described by ${\psi}_a(m)$ which obey the usual equal-time commutation relations

\begin{equation}
\left[{\psi}_i(\vec{r}),{\psi}_j(\vec{r}')\right]=0,\left[{\psi}_i(\vec{r}),{\psi}^{\dagger} _j(\vec{r}')\right]=\delta_{ij}\delta(\vec{r}-\vec{r}')\ .
\end{equation}

One obtains the linearized Hamiltonian, by discarding terms beyond the second order in the fluctuations. The contribution of the fluctuation part to the above Hamiltonian is given by,

\begin{equation}
K=\sum_{i,j=1} ^4 \int V_i ^\dagger M_{ij} V_j\ ,
\end{equation}

where $(V_1,V_2,V_3,V_4) \equiv (\psi_a,\psi_m,\psi_a ^\dagger,\psi_m ^\dagger)$ and

\begin{widetext}
$M=\left[\begin{array}{cccc}
-\frac{1}{2}\partial_{x}^{2}+\lambda_{am}N|\phi_{m}|^{2} & \lambda_{am}N\phi_{a}\phi_{m}^{*}+\alpha\sqrt{2N}\phi_{a}^{*} & \frac{1}{2}\lambda_{a}N\phi_{a}^{2}+\alpha\sqrt{\frac{N}{2}}\phi_{m} & \lambda_{am}N\phi_{a}\phi_{m}\\
\lambda_{am}N\phi_{a}^{*}\phi_{m}+\alpha\sqrt{2N}\phi_{a} & -\frac{1}{4}\partial_{x}^{2}+\epsilon+\lambda_{am}N|\phi_{a}|^{2} & 0 & \frac{1}{2}\lambda_{m}N\phi_{m}^{2}\\
\frac{1}{2}\lambda_{a}N\phi_{a}^{*2}+\alpha\sqrt{\frac{N}{2}}\phi_{m}^{*} & 0 & -\frac{1}{2}\partial_{x}^{2}+2\lambda_{a}N|\phi_{a}|^{2} & 0\\
\lambda_{am}N\phi_{a}^{*}\phi_{m}^{*} & \frac{1}{2}\lambda_{m}N\phi_{m}^{*2} & \frac{1}{2}\lambda_{m}N\phi_{m}^{*2} & -\frac{1}{4}\partial_{x}^{2}+2\lambda_{m}N|\phi_{m}|^{2}\end{array}\right]$
\end{widetext}

Taking cue from \cite{eberlystability}, the AMBEC system is stable if all the eigenvalues of M are non-negative, i.e., if M is semi-positive. The system is unstable if the lowest eigenvalue of M is negative. This is justified by the fact that arbitrary fluctuations increasing the energy of the system, should mean that the system was stable in the first place. Similarly, if the fluctuations decrease the energy of the system, the system was not stable to start with. The analysis assumes nothing regarding the nature of the fluctuations. The lowest eigenvalue determines the stability of the mean fields. It has been pointed out that the fluctuations can also be in particle numbers and that the consideration of these fluctuations would provide a way to probe the effective interactions between particles. 

Evolution of the mean fields is governed by the NLSE and small perturbations of the mean fields will remain bounded if all the normal mode frequencies (or collective excitation frequencies) of the linearized system are real. The collective excitation frequencies are defined by

\begin{equation}
\eta M\left[\begin{array}{c}
u_{1k}\\
u_{2k}\\
u_{3k}\\
u_{4k}\end{array}\right]=\omega_{k}\left[\begin{array}{c}
u_{1k}\\
u_{2k}\\
u_{3k}\\
u_{4k}\end{array}\right]
\end{equation}

where,

\begin{equation}
\eta=\left[\begin{array}{cccc}
1 & 0 & 0 & 0\\
0 & 1 & 0 & 0\\
0 & 0 & -1 & 0\\
0 & 0 & 0 & -1\end{array}\right]
\end{equation}

with $(u_{1k},u_{2k},v_{1k},v_{2k})$ being the mode functions. A semipositive M does indeed guarantee stability of the mean fields. A necessary but not sufficient condition for mean fields to be unstable (have complex normal mode frequencies), is that the lowest eigenvalue of M be negative \cite{Blaizot}. The final task is to compute the lowest eigenvalue of M, for varying values of equation parameters. It is done by discretizing in space. We fixed time to be 0 and assumed that the eigenfunctions of the matrix vanish at $\pm \infty$. We varied $\lambda_{am}$, the coupling constant for the interaction between atomic and molecular species, the particle number N and the interconversion coupling constant $\alpha$. 


\begin{figure}[t]\noindent \centering{} \includegraphics[scale=0.4]{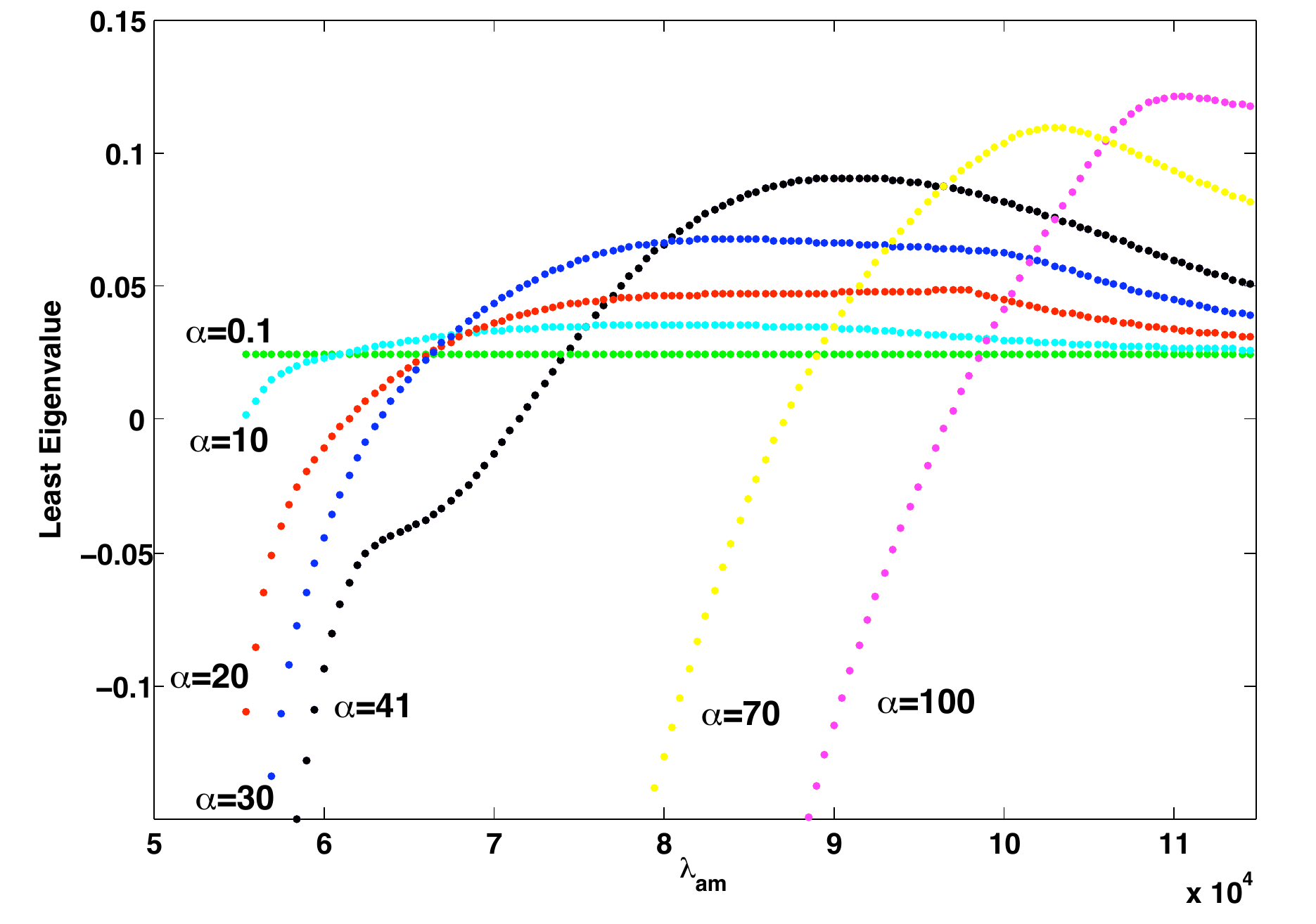}\caption{Stable parameter domain under quantum fluctuations to the coupled soliton solutions of the bright-grey type. $\lambda_{a}=3$x$10^{4}$, $\lambda_{m}=10^{5}$, $\epsilon=0$ and N=100. Varying the particle number (N) does not change the plots by a considerable extent. The curves are very sensitive to the interconversion coefficient $\alpha$. The `stable-unstable' transition occurs at different values of $\lambda_{am}$ for different $\alpha$. } \end{figure}

From this analysis, we have identified a region in the parameter space where the solution makes a transition from `unstable' to `stable'. Fig (3) shows the least eigenvalue of M, computed as a function of atom-molecule interaction coefficient $\lambda_{am}$. Different curves were obtained from using different values of the interconversion coefficient $\alpha$. We then use the stable parameter domain in simulating our solution. 

\section{Numerical Study of soliton solutions}
We have further studied the evolution of the bright-grey soliton solution numerically. The following are the results of the simulations carried out on the bright-grey soliton pair. Figs 4 and 5 were obtained by the stable finite difference scheme, Crank Nicolson (CN). It is an implicit scheme and accumulates second order error in both space and time steps.

\begin{figure}[t]
\noindent \centering{}
\includegraphics[scale=0.4]{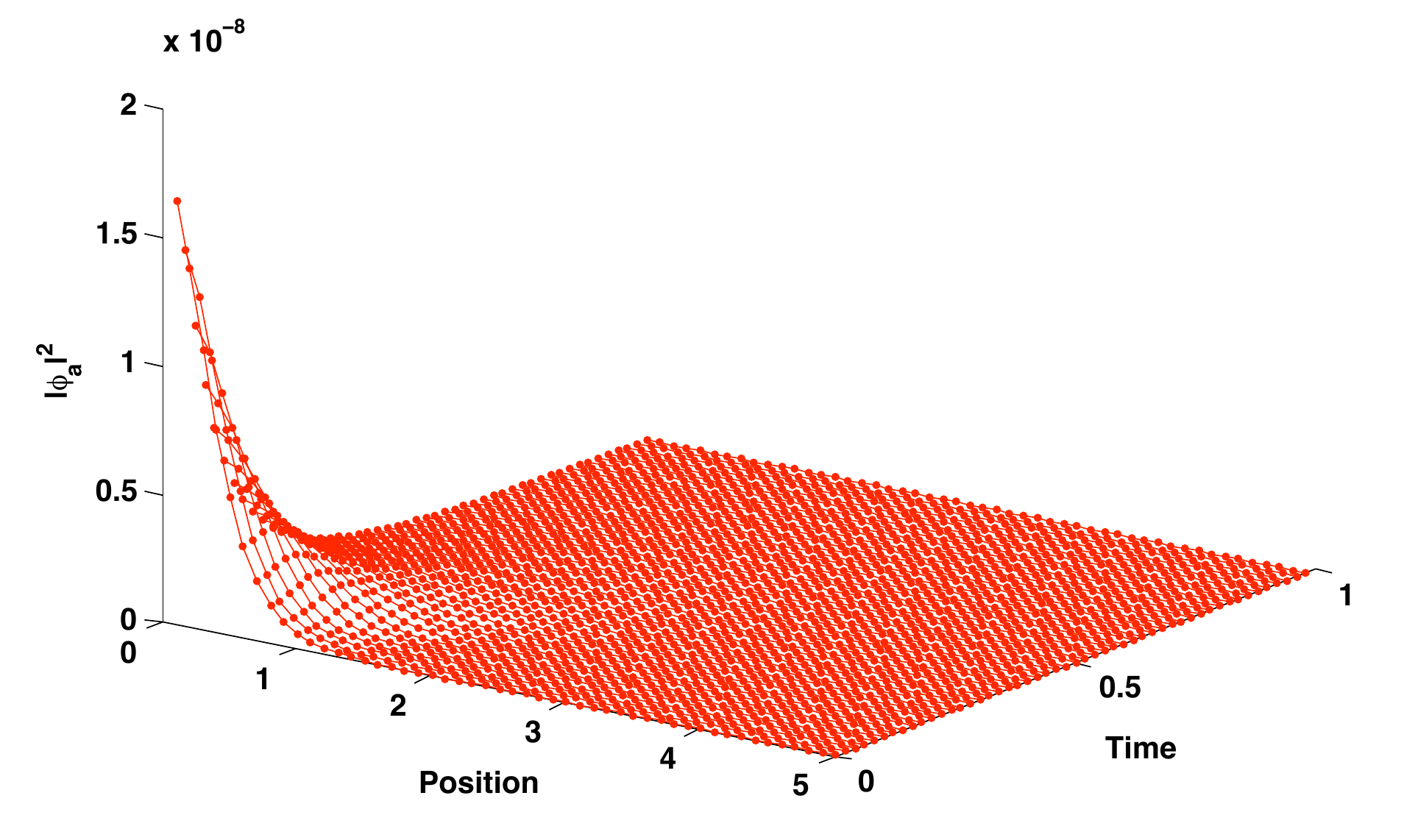} \includegraphics[scale=0.4]{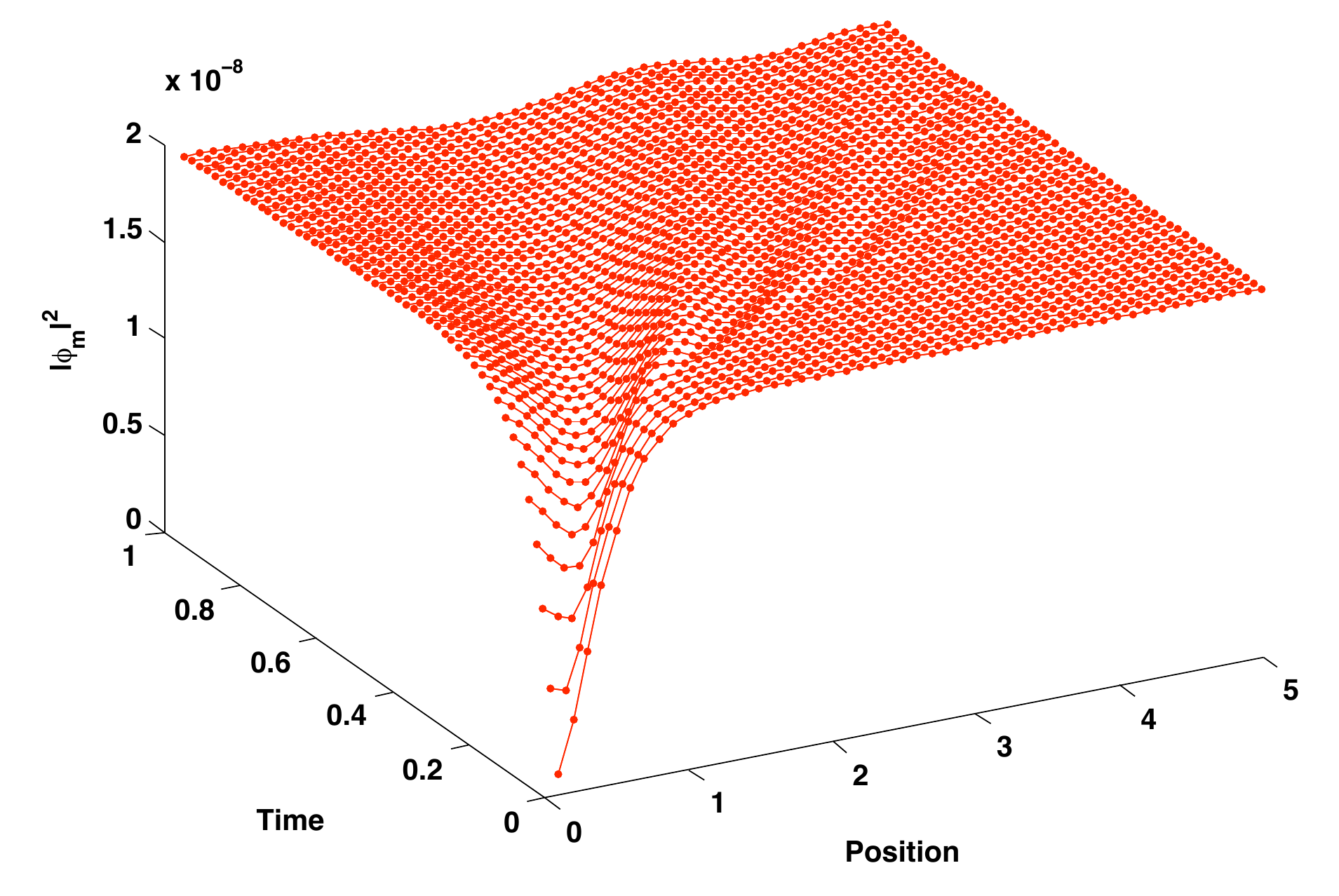}
\caption{Results of numerical evolution of the bright and grey pair solitons in the stable domain. The parameters used were obtained from the stable domain of $\lambda_{am}$ in Fig (3), and are  $\lambda_a=30\mathrm{x}10^3$, $\lambda_m=10^5$,  $\lambda_{am}=80.5\mathrm{x}10^3$,   $\alpha=20$,   N$=100$, and $\epsilon=0$. The least eigenvalue in this case is $+0.0466$, indicating stable domain.}
\end{figure}

\begin{figure}[t]
\noindent \centering{}
\includegraphics[scale=0.4]{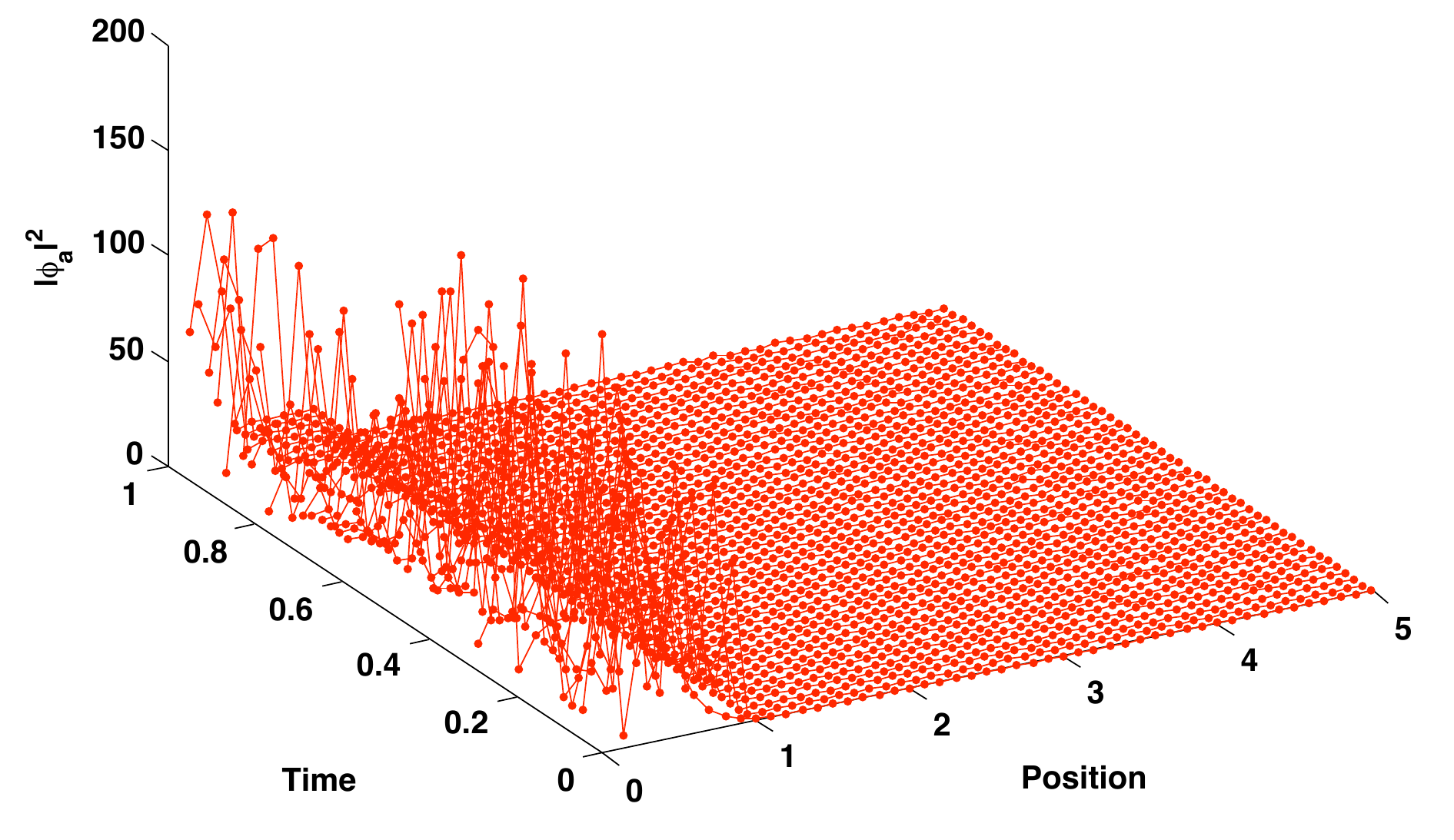}
\includegraphics[scale=0.4]{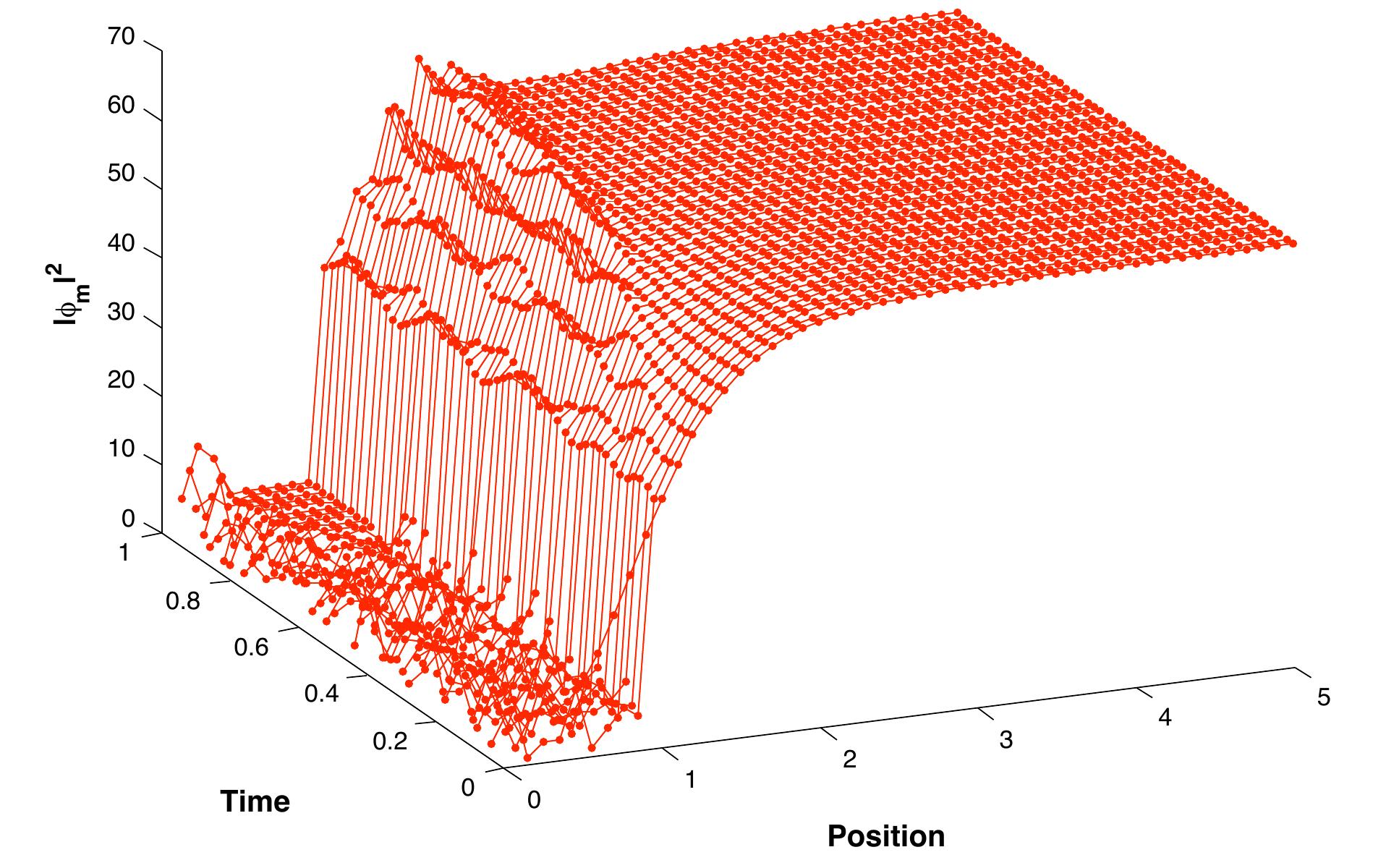}
\caption{Results of numerical evolution of the bright and grey pair solitons in the unstable domain. The parameters used are  $\lambda=0.15$, $\lambda_m=0.50$,  $\lambda_{am}=0.506$,   $\alpha=41.5$,   N=$1000$, and $\epsilon=-900$. The least eigenvalue in this case is $-4475.8$ indicating unstable domain.}
\end{figure}

The following simulations were implemented using the coupled split-step and CN scheme \cite{adhikari}. The solution is first evolved using split-step method, using only the terms containing atom-atom, molecule-molecule and atom-molecule interactions and the binding energy. The CN algorithm next evolves this evolved part, using the atom-molecule interconversion and dispersion terms. The results are shown in the following figures.

\begin{figure}[t]
\noindent \centering{}
\includegraphics[scale=0.4]{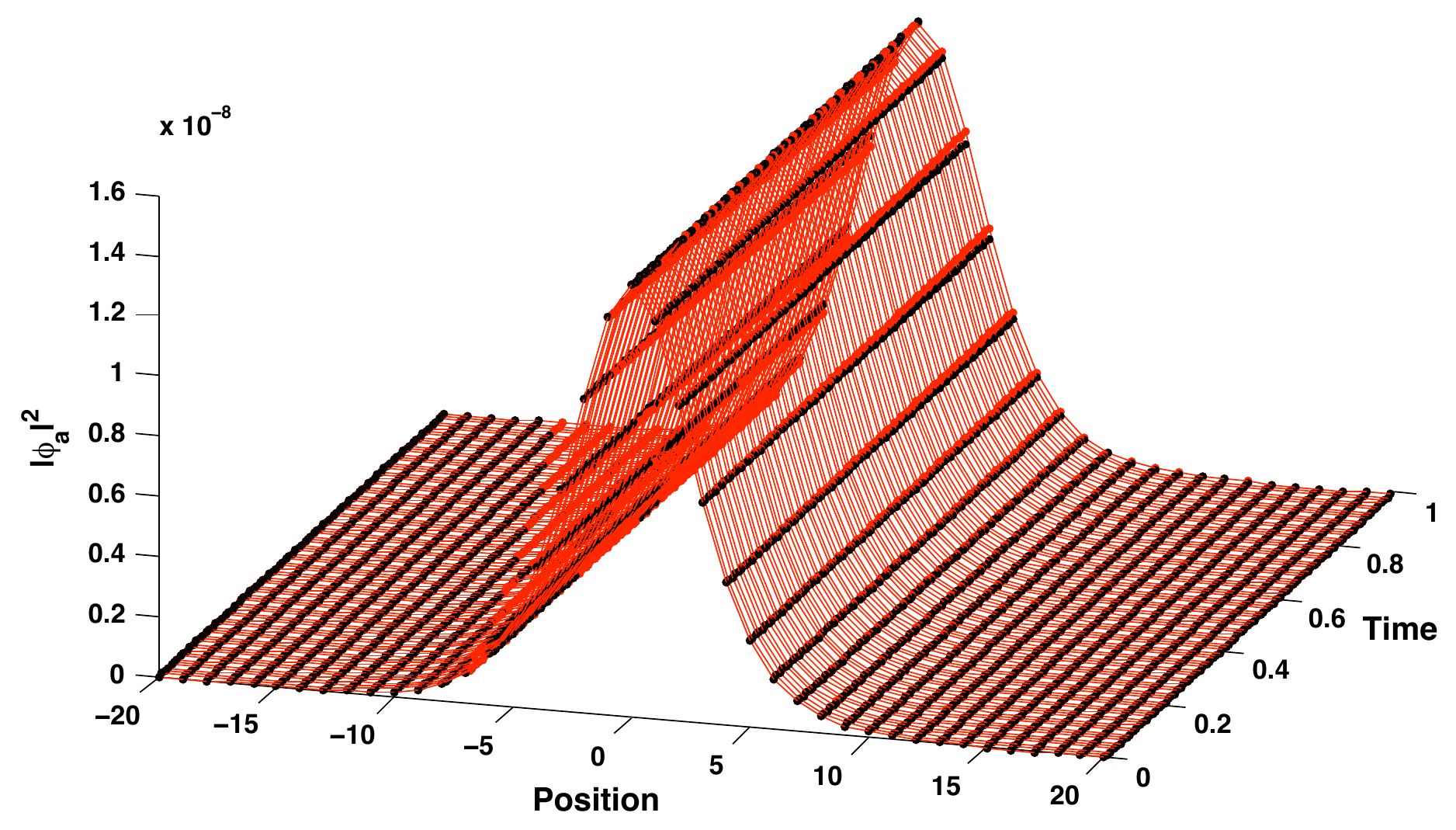}
\includegraphics[scale=0.4]{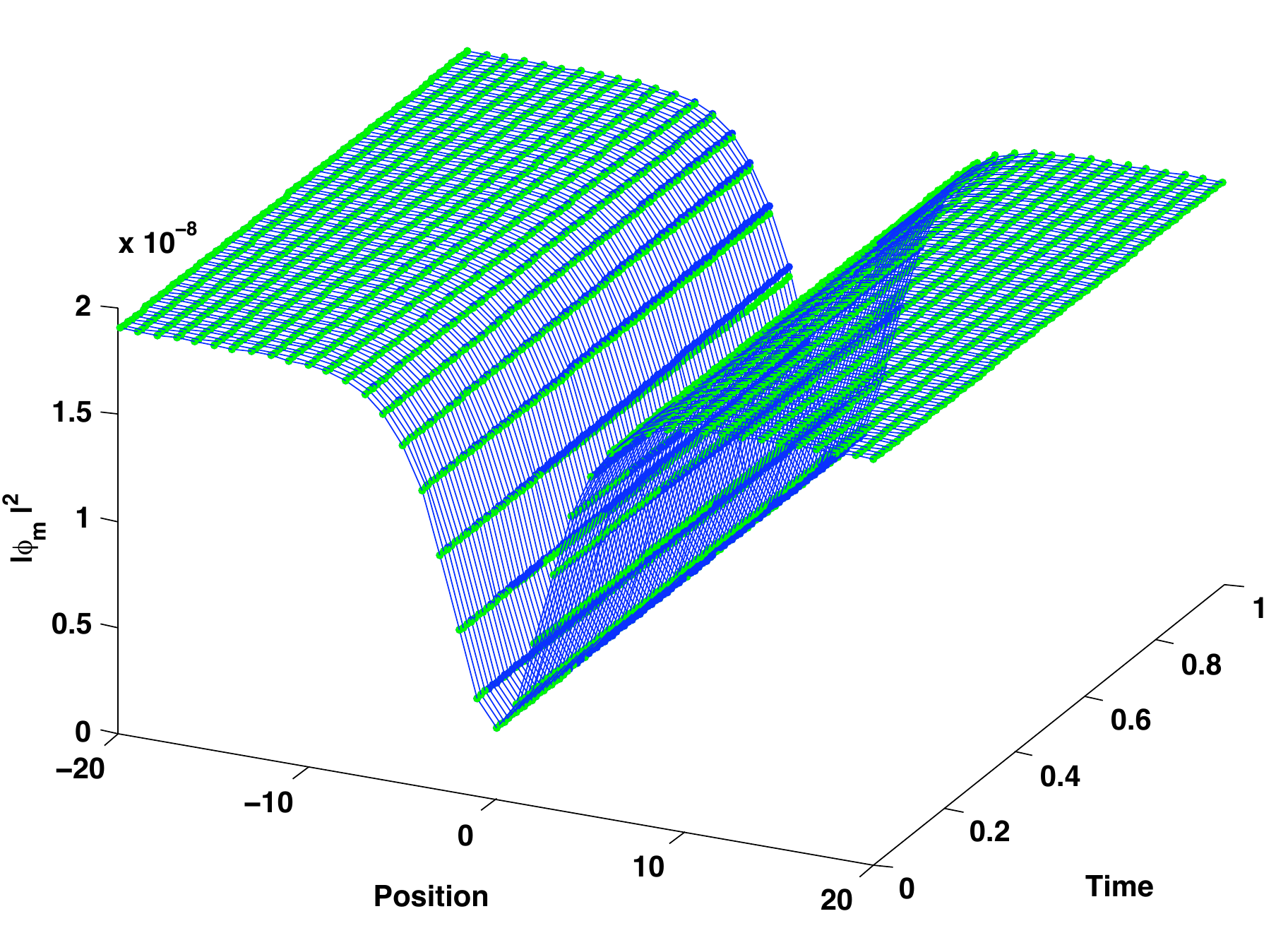}
\caption{Results of numerical evolution of the bright and grey pair solitons in the stable domain. The parameters used were obtained from Fig 3 and are  $\lambda_a=30\mathrm{x}10^3$, $\lambda_m=10^5$,  $\lambda_{am}=80.5\mathrm{x}10^3$,   $\alpha=20$,   N$=100$, and $\epsilon=0$. Red and blue denote the numerically evolved solutions, black and green, denote analytical solutions (5) and (6). A good match between the two indicates that the solutions in this domain are stable.}
\end{figure}

The results depicted in Figs 4 and 5 suggest that the solutions are stable when the parameters are chosen from the stable region of Fig 3 and unstable when the least eigenvalue is highly negative. However, an exhaustive study needs to be done to conclude if the analysis performed on our solution correctly predicts stable-unstable domains. It would also be worthwhile to see what predictions linear stability analysis will have for stable-unstable domains and whether or not there is any overlap in the domains predicted by these two analyses.

\section{Modulation Instability}

We now proceed to study the possibility of modulational instability in this system. In NLSE, the standard way in which bright solitons and solitary wave structures are generated, is through modulation instability (MI). In this case, the continuous wave solution becomes unstable. MI of a nonuniform initial state in the presence of a harmonic potential has been studied both analytically and numerically in the context of the mean field of the BEC \cite{Carr}. The analysis of MI in AMBEC is similar to that in the two component BEC (TBEC), but not exactly the same \cite{rajuppporsezian}. 

\subsection{Gain Spectrum}

To find out the domain of MI in any system, in general one proceeds as follows. We first find out continuous wave solutions to the coupled GP equations that are fixed in space. Subsequently one applies space-time dependent perturbation to this solution and finally, the gain spectrum is obtained. For this purpose, we use the ansatz

\begin{align}
\phi_{a}(x,t)&=(\phi_{a0}+\epsilon_{a}(x,t))\exp(i\psi_{a}t)\ ,\\
\intertext{and} \phi_{m}(x,t)&=(\phi_{m0}+\epsilon_{m}(x,t))\exp(i\psi_{m}t)\ .
\end{align}

Then we assume, 

\begin{align}
\epsilon_{a}(x,t)&=u_{a}\exp[i(kx-\Omega t)]+v_{a}^{*}\exp[-i(kx-\Omega t)]\\
\epsilon_{m}(x,t)&=u_{m}\exp[i(kx-\Omega t)]+v_{m}^{*}\exp[-i(kx-\Omega t)]
\end{align}

where $u_{a(m)}$ and $v_{a(m)}^{*}$ are to be determined. The following represents the consistency condition, that (32)-(35) yield valid solutions to the coupled GP equations, in terms of a matrix determinant:

\begin{widetext}
\begin{align}
\left|\begin{array}{cccc}\gamma_1(k) & -\Omega & 2\lambda_{am}N\phi_{a0}\phi_{m0}+\alpha\sqrt{2N}\phi_{a0} & 0\\\Omega & \gamma_2(k) & 0 & \alpha\sqrt{2N}\phi_{a0}\\2\lambda_{am}N\phi_{a0}\phi_{m0}+\alpha\sqrt{2N}\phi_{a0} & 0 & \gamma_3(k) & -\Omega\\0 & \alpha\sqrt{2N}\phi_{a0} & \Omega & \gamma_4(k)\end{array}\right|=0
\end{align}
\end{widetext}

where

\begin{align}
\gamma_{1}&=k^{2}/2+3\lambda_{a}N\phi_{a0}^{2}+\lambda_{am}N\phi_{m0}^{2}+\alpha\sqrt{2N}\phi_{m0}\\
\gamma_{2}&=k^{2}/2+\lambda_{a}N\phi_{a0}^{2}+\lambda_{am}N\phi_{m0}^{2}-\alpha\sqrt{2N}\phi_{m0}\\
\gamma_{3}&=k^{2}/4+\epsilon+3\lambda_{m}N\phi_{m0}^{2}+\lambda_{am}N\phi_{a0}^{2}\\
\gamma_{4}&=k^{2}/4+\epsilon+\lambda_{m}N\phi_{m0}^{2}+\lambda_{am}N\phi_{a0}^{2}
\end{align}

One obtains a quadratic equation in $\Omega^{2}$, which gives two roots. MI sets in when $\Omega^{2}<0$. The growth rate is given by the imaginary part of $\Omega$. Plotting this as a function of varying parameters, one gets the gain spectrum. We study one of the branches ($\Omega^+$ branch), corresponding to one of the roots of the above mentioned quadratic equation. The other branch $\Omega^-$ was also studied and yielded a similar gain spectrum.

\begin{figure}[t] \noindent \centering{}\includegraphics[scale=0.4]{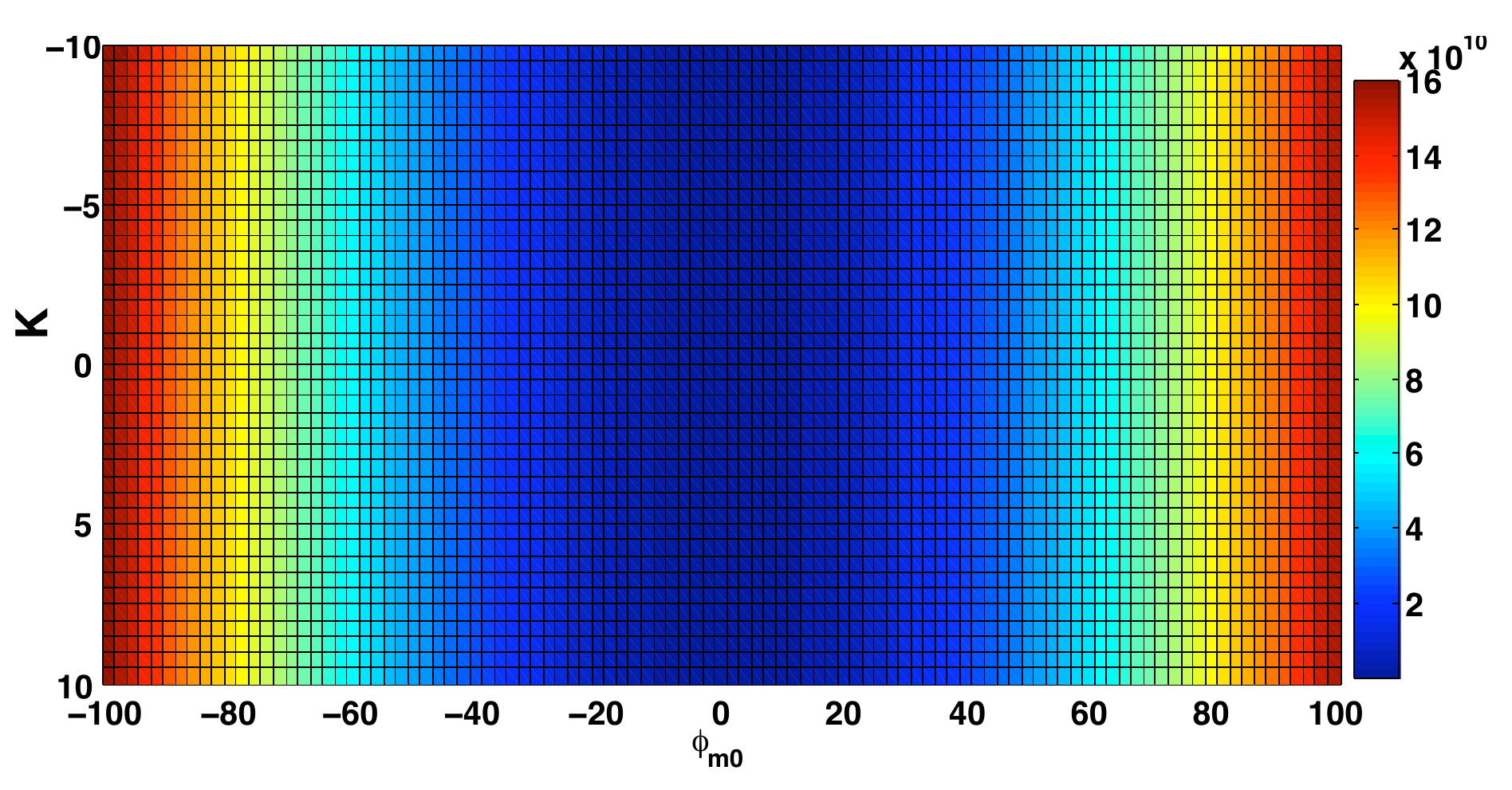}\caption{Gain Spectrum for MI in AMBEC. The following are the values of the parameters that were used: 
 $\lambda_a =30\mathrm{x}10^3$, $\lambda_m=10^5$, $\lambda_{am}=90\mathrm{x}10^3$, $\alpha=0$, N $=100$, $\epsilon=0$. } \end{figure}
 
Fig 7 gives the gain spectrum for MI in the AMBEC system. The two modulationally unstable species, atoms and molecules, may appear as propagating periodic or localized solitary waves. 

\section{Conclusion}
In conclusion, we have found new cross-phase modulated localized soliton solutions for an AMBEC. Owing to the difference in nature of the atom-molecule interconversion terms in Eqs (1) and (2) and the equivalent terms in a TBEC, the solutions vary significantly in these two cases. Many of the solutions valid for the case of a TBEC, do not obey Eqs (1) and (2). We have identified three solutions for the case of AMBEC and have devised a mechanism for obtaining more. We have analyzed one of these solutions, the bright-grey pair,  for stability under quantum fluctuations, by performing an analysis presented in \cite{eberlystability}. The analysis helped us predict stable and unstable regions in the parameter space. This was supported by the numerical simulations (Figs. 4-6). We also obtained the domain of modulation instability in the AMBEC system.

\end{document}